# FakeCovid- A Multilingual Cross-domain Fact Check News Dataset for COVID-19

**Gautam Kishore Shahi** [1] **and Durgesh Nandini**[2]
University of Duisburg-Essen, Germany[1], University of Bamberg, Germany[2]
gautamshahi16@gmail.com [1] durgeshnandini16@yahoo.in[2]

## Abstract

In this paper, we present a first multilingual cross-domain dataset of 5182 fact-checked news articles for COVID-19, collected from 04/01/2020 to 15/05/2020. We have collected the fact-checked articles from 92 different fact-checking website after obtaining references from Poynter and Snopes. We have manually annotated articles into 11 different categories of the fact-checked news according to their content. The dataset is in 40 languages from 105 countries. We have built a classifier to detect fake news and present results for the automatic fake news detection and its class. Our model achieves an F1 score of 0.76 to detect the false class and other fact check articles. The FakeCovid dataset is available at Github.[1]

## Introduction

Coronavirus disease 2019 (COVID-19) is a contagious disease caused by SARS coronavirus 2, a virus closely related to the SARS virus(Moriguchi et al. 2020). The first case of the virus was discovered in the city of Wuhan (Cascella et al. 2020) in late December 2019. From the beginning of March, it was recognised to be a global issue and was later declared a pandemic by WHO. According to WHO, COVID-19 has spread to at least 215 countries around the globe (Organization and others 2020a). People from all over the world are using keywords such as covid19, coronavirus etc. for discussion. So, perhaps for the first time in history, we see that humanity has been exposed to substantial interaction on a single subject in dozens of different languages and on many platforms. In parallel, the "infodemic" of rumours and misinformation related to the virus came to surface.

Infodemic is a term coined by World Health Organization (WHO) to explain the misinformation of virus, and it makes difficult for users to find reliable sources for any claim made on the pandemic, either on the news or social media (Organization and others 2020b; Zarocostas 2020). Hence, browsing or gathering information from the news media or social media platforms without checking the correctness affects people's psychology, daily lives, and behaviours. On the other hand, misinformation is a piece of false information or inaccurate information. It could be a false rumour, wrong claim, insult and prank. Around the globe, scientists have been trying their best to discover remedies for the COVID 19, and the infodemic that scares the planet by words have become manifolds threatening. The fight against the pandemic not only involves finding remedies against the virus but it is also crucial to delicately deal with "infodemic" of misinformation so that the authenticity related to the disease and the psychological health of people around the globe is reassured.

Many people are involved in the creation and consumption of misinformation related to COVID-19. Old messages and incidents are repurposed to spread fake news by somehow connecting them to the subject of the virus. The government of various countries and mediators of social media platforms have been consistently trying to stop the publishing of fake news but the process has been ineffective due to numerous hurdles (Tidy 2020). One of the key issues with detecting misinformation related to COVID-19 is that there is a lack of corpus to test methods for fake news detection.

In this paper, we have presented a human-annotated multilingual cross-domain fact-checked set for the COVID-19. According to the Google Trends from 1st January 2020 to 15th May 2020, the term fake news and coronavirus have been repeatedly searched in last three months. The search rate was maximum during March 2020 as shown in figure 1 and 2. Both trends follow the same curve during the March and April, which indicates people are eager to know or check the truth of coronavirus(COVID-19).

Fake news detection is a cumbersome task for two reasons: firstly, it requires a specially trained human to make a clear distinction between fake news and real news, and secondly, a huge velocity, veracity and diversity of fake news are available on various social media platforms, newspapers, and news channels in multiple domains. Due to the lack of efficient skills and human resources, an automatic tool for fake news detection is required. Variety and velocity of fake news keep changing, so the existing methods fail to detect misinformation consistently.

The main contribution of this work is to prepare an open-source data set for detection of misinformation at the time of the pandemic. A machine learning-based classifier is built to detect the misinformation at the time of the pandemic.

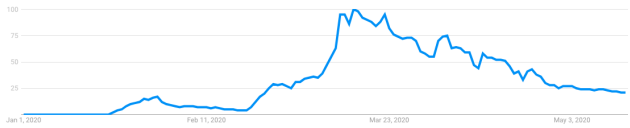

Figure 1: Google Trend on Coronavirus

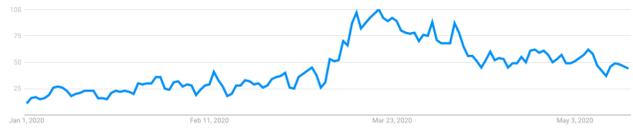

Figure 2: Google Trend on Fake News

In the following sections, we discuss the related work, data collection, data annotation, data cleaning and prepossessing section, data exploration, classification results, discussion and finally we discuss the conclusion and ideas for future works.

## Literature

To examine the fact check we need a reliable corpus to run our model. From last few years, a collection of small datasets related to fact-checking has been published. These datasets are a mixture of different topics of fake news, For instance, US election 2016(Wang 2017). Most of them are only in the English language and have been collected from only a few sources which limit the diversity of fact-checking.

(Augenstein et al. 2019) describes the different kinds of the fact-checking dataset available and their limitations. The authors have come up with a multi-domain, evidence-based fact-checking dataset. They have also described the metadata of Fact-check articles. (Vlachos and Riedel 2014) describes the task of fact-checking and construction of dataset using the process used by journalists. Several methods are published for automatic detection of fake news, (Pérez-Rosas et al. 2017) discuss the automatic detection of fake news and linguists difference in false and legitimate content. (Zhou and Zafarani 2018) analyse, compare and summarise the several different methods available for fake news detection. Authors give an overview of methods which have been tested for fake news detection. Research has been done to look inside on the feature of a news story in the modern diaspora along with different kind of news story and its impact on people(Parikh and Atrey 2018).

Both COVID-19 pandemic and infodemic spread in parallel. (Mesquita et al. 2020) proposes a framework to fight against fake medical news because fake news can intensify the effect of COVID-19 pandemic. All health worker, scientists, the government are trying to fight against the fake medical news. During March 2020, several cases have been discovered where people are consuming the partially true information about COVID-19 on Facebook and WhatsApp without using the proper medical terms, which creates a panic about medicine and prevention for COVID-19 (Orso et al. 2020). By analysing the search behaviour of people in Italy from January to March 2020 and found that a "large number of infodemic monikers were observed across Italy" (Rovetta and Bhagavathula 2020).

During the time of the pandemic, fake news is spread all over the world in different languages. The fact-news is covered in different domains like origin and spread, conspiracy theory etc. There is a lack of resources which are multilingual and cross-domain and have been collected from multiple sources.

**Contributions**: To solve the problems of a multi-purpose fact-checked corpus, we provide an open-source dataset which is multilingual, cross-domain and collected from 105 countries. The corpus can be used for the several studies related to COVID-19 and fake news.

## Data Collection

This section explains the steps followed to collect the data from different fact-checking websites. We have used the following data source for taking a reference to the fact checking website.

**Snopes**- Snopes (Snopes 2020) is an independent publication owned by Snopes Media Group. Snopes verifies the correctness of misinformation spread across several topics. As for the fact-checking process, they manually verify the authenticity of the news article and performs a contextual analysis. In response to the COVID-19 infodemic, Snopes provides a collection of a fact-checked news article in different categories based on the topic of the article.

**Poynter**- Poynter (Poynter Institute 2020) is a non-profit making institute of journalists. In COVID-19 crisis, Poynter came forward to inform and educate to avoid the circulation of the fake news. Poynter maintains an International Fact-Checking Network(IFCN), the institute also started a hashtag #CoronaVirusFacts and #DatosCoronaVirus to gather the misinformation about COVID-19. Poynter maintains a database which collects fact-checked news from 91 fact-checking organization in 40 languages.

For data collection, we used the Poynter and Snopes as a bridge to get the links to the original fact-checking website. The steps involved in the process of data collection:

**Fact-check articles** We weekly scraped the list of fact-checked news mentioned at reference sources (Poynter & Snopes) and collected the list of the fact-checked news articles mentioned on their website. We have assigned a unique identifier to each of them and its denoted by FCID.

**Source of articles** From the list of fact-checked news articles, we fetched the reference to the source of the fact-checked article which was published by the Poynter and Snopes. This is represented by article_source.

**Title** From the collected link, we fetched the tile of the news article. In many cases, the title provided by the reference provided(Poynter & Snopes) is different from the original title given by the fact-checking website. Title provided by the reference provide is defined by reference title and denoted by ref_title. Similarly, source title is defined source title and represented as source_title.

**Published Date** We collected the published date, which refers to the date of publication of fact-checked articles by the fact-checking websites. It is represented by published_date.

**Content of articles** Once we have a source of a news article, then we used Beautiful shop (Richardson 2007), python-based library to crawl the HTML Document Object Model(DOM) to gather other information like textual content, country, date of the article. It is defined by source_title.

**Class** The fact checking website assign a class for each fact checked article. Each fact checking website have a set of classes designed by them, for instance the classes may be 'true', 'false', 'partially false' meaning that the article is true, false or partially false respectively. The articles are then elucidated by them into one of their designed classes. A detailed description is mentioned in table 6. This attribute is denoted by class.

**Social media link** Fake news is primarily circulated over social media to be spread among a large number of people, and the fact-checker provides a reference to social media posts, so we have collected the presence of social media on the fact-checked articles. Once we extracted the content from the web page, we looked for the social media presence in the news article. To check the twitter presence, we look for keyword twitter and status in the hyperlink. For YouTube, we look for the keywords "youtube" and "watch". For Reddit, we look for the keyword "Reddit" and the pattern "reddit.com/r/" using regex. Similarly we check the presence of Facebook and Instagram using regex pattern. This property is denoted by SM_Link.

**Fact checking website** From the fact-checked references, we also fetch the source of the article, which is the fact-checking company. This property is denoted by verifiedby.

**Country** We fetched the country of the articles, which tells in which country the article was circulated. Sometimes the same articles were circulated in several countries. In our dataset, the maximum number of countries that an article appeared to be circulated around was four, hence we divided the country attribute into 4 sub attributes, namely country1, country2, country3 and country4 to identify each country separately.

**Category** Fake news are circulated regarding several topics for instance, origin or virus, or international response. The category attribute has been used to denote under which topic the article falls into. The value of the attribute has been assigned by us after manually annotating the articles. A detailed description of the annotations is mentioned in section "Data Annotation".

**Language** The fact check articles are circulated in different countries in several languages. To identify the language of the article we created the attribute lang which denotes the language of the fact checked articles. The process of language detection is discussed in section "Data cleaning and Prepossessing".

We have collected 5182 articles circulated in 105 countries from 92 fact-checkers. Articles published were from 04.01.2020 to 15.05.2020 in 40 languages among which 40.8 % articles are in the English language. A summary of the dataset is presented in table 1.

Table 1: Data Description.

| Element | Count |
|---|---|
| Fact checked article | 5182 |
| class of fact check article | 23 |
| Fact check Category | 11 |
| Fact checking website | 92 |
| Country | 105 |
| Language | 40 |

## Data Annotation

In the context of COVID-19, there are variants(category) of fake news circulating. So we annotated the articles for each category. For annotation, we defined the task to label the news article into a predefined category according to the content of the news article. After doing the content analysis, we decided the category of the news article.

Annotators followed a standard procedure for annotation, visit the source article using the link forward and look at the title and read the content of the article then decided the category of the article. We have categorised the news articles into 11 distinct categories based on their contents, and the categories were also reviewed by professional fact-checker. The description of the categories are mentioned in table 7. Due to our limitation with the knowledge of the language we have annotated 1951 fact check articles, in three language English(2116), Hindi(141) and German(47).

**Annotation** For annotation of news articles, we selected three people, based on background knowledge and linguistic knowledge. Data is being annotated by one annotator and the second annotator annotates the randomly chosen news article according to the language to calculate the intercoder reliability (Lombard, Snyder-Duch, and Bracken 2002) agreement.

The first annotator is pursuing a master's degree in our chair, and she has a good experience of working with data annotation. The annotation guidelines were shared in the beginning and the student was asked to provide sample annotation, after verifying the annotation quality, the annotator was asked to label other data, as a fluent English speaker and native German speaker, she was asked to annotate English and German news article. The second annotator is a PhD student who is an experienced data scientist. He is a fluent German Speaker and native Hindi speaker. He annotated all the Hindi news articles and some of the randomly chosen German news article. The third annotator, a PhD student and working in the area of machine learning, is fluent in English and is a native Hindi speaker.

The annotator annotated randomly chosen English and Hindi news article. For each language, we calculated reliability score to measure the agreement between two annotators. The number of articles labelled by all three annotators and their intercoder reliability is shown in table 2.

Table 2: Inter coder reliability

| Annotator | English | Hindi | German |
|---|---|---|---|
| 1st | 2116 | -- | 47 |
| 2nd | -- | 114 | 23 |
| 3rd | 100 | 46 | -- |
| **Inter-coder reliability** | 91% | 96% | 94% |

## Data Cleaning and Prepossessing

### Data Cleaning

We have used the following cleaning process to remove unwanted information from collected data.
**Faulty URLs** Some of the URLs are entered wrong at the fact-checking website while some do not exist anymore. So we manually analysed the faulty URLs and either corrected them if they were wrongly entered or removed them if they ceased to exist.
**Missing title** In some cases the title of the article is missing so we manually added source title after looking into the webpage of the article.
**Removing duplicates** Some of the fact-checked articles were duplicate so we filtered the unique article using the URL of the article and removed the duplicates.

### Data Prepossessing

In this step, we applied the basic Natural Language Processing(NLP) preprocessing techniques to remove unwanted information from the data like lowercase, removal of the short word, tokenization etc. For data cleaning we have used python library NLTK(Loper and Bird 2002), Textblob(Loria 2018) and regular expression.
**Language detection** The data collected were in multiple languages. To identify the language of the content, we performed language detection using langdetect, a python based library (Shuyo 2014). We used the content of the articles to assign them their respective language.
**Abbreviations and contractions** Abbreviations and contractions of words both serve to shorten a word, but while abbreviations omit the last few letters of the word, contractions omit letters in the middle of the word. Abbreviations and contractions have become even more common with the internet, texting, and the need to keep messaging and posting text to a minimum.
**Spelling correction** Sometimes, tweets may contain typo errors. We used Textblob, a python library for text processing. Textblob returns the corrected words as output if any sentence has spelling mistakes.

## Data Exploration

In this section, we performed the exploratory analysis of the dataset.

**Across country** We have analysed the circulation of fact check data across different countries. We found that India has a maximum number of cases of fake news, followed by the USA and Spain. The count of fact check articles collected from the top 10 countries is shown in table 3.

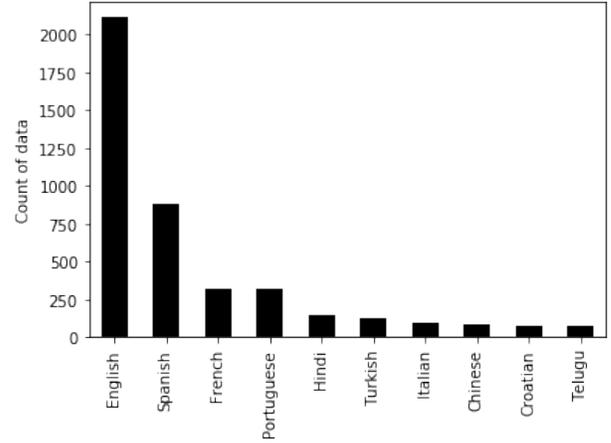

Figure 3: The language(top 10) distribution prior of fact check article

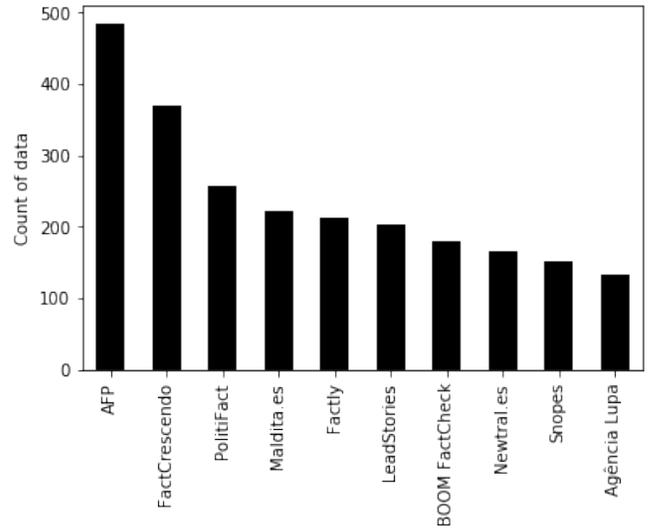

Figure 4: Distribution of Top 10 Fact-checking website

Table 3: Count of fact-chek articles in top 10 country

| Country | Count |
|---|---|
| India | 1083 |
| United States | 677 |
| Spain | 426 |
| Brazil | 283 |
| France | 229 |
| Philippines | 157 |
| Columbia | 151 |
| Taiwan | 132 |
| Turkey | 126 |
| Italy | 111 |

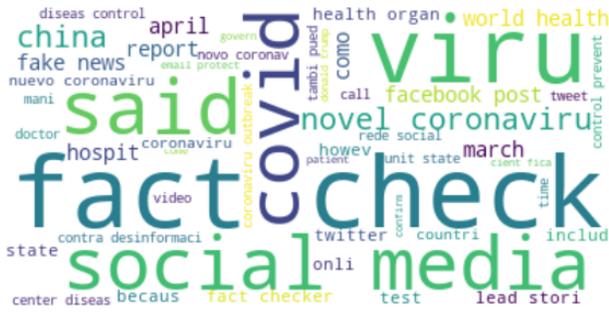

Figure 5: Top 100 Word from the data set

**Across language** As the fake news spread across all over the globe, it also circulated in the different national or regional language of the respective countries. We filtered the top 10 languages as per the count of the fact check articles and result is presented in figure 3.

**Word distribution** To have an overview of the idea for the content of the news article. We have picked the top 100 words and have plotted the word cloud as shown in figure 5. The figure shows that the words covid, fact check, social media are dominant.

**Fact checking website** We have presented the dataset from 105 different countries and each fact checking website works in a particular region and language. So we have plotted the number of articles published by top 10 fact checking website in figure 4. AFP[2] covers maximum number of fact checked articles for the COVID-19.

**Social media link** The fact-checked articles embed posts from social media in their articles to discuss their alleged claims. We collected the news articles which have some social media link embedded in it. Count for the presence of different social media link is presented in table 4.

Table 4: Count of social media embedded link.

| SM Platform | Count | Percent |
|---|---|---|
| Twitter | 916 | 17.7% |
| Facebook | 1990 | 38.4% |
| Reddit | 106 | 2.05% |
| YouTube | 602 | 11.6 % |
| Instagram | 193 | 3.72 % |

## Classification and Results

We filtered the data using classes mentioned in table 5. We divided the data set into two class, false and others. False includes the articles which are labelled false by fact-checker. There are 4132 articles in the false category. Apart from the false class, the remaining 22 class are considered as other categories, which counts 1050 facts checked articles. We decided to go for two classes because 78 % data is in the false category, so if we can filter the false category from a stream of misinformation, it will ease the job of early-stage screening for fact-checker. We tested our model for the English language and the results are: the total number of false and other class are 2116 and 500 respectively.

---

[2]https://factcheck.afp.com/

We have used a BERT based classification model without fine-tuning as a single view to measure the performance of the algorithm. We set hidden units as 300 and training epoch as 150. Each training process continues until the restriction or validation loss is continued. Batch size is set to 1, and the learning rate is 0.001. The classification result obtained is shown in table 5.

Table 5: Classification Results

| Class | Precision | Recall | F1-Score |
|---|---|---|---|
| False | 0.58 | 0.75 | 0.65 |
| Others | 0.87 | 0.75 | 0.80 |
| Overall | 0.78 | 0.75 | 0.76 |

## Discussion

Gathering the fact check news article is a critical and challenging task due to reasons such as unavailability of references, and presence of false references. For some of the fact-checked article, the reference links are not available or invalid so we could not collect details content_text, ref_title. This could be a mistake while aggregating the dataset. Poynter has multiple duplicates of news articles, which filtered the fact-checked news article using ref_url but there might be some other duplicates.

Some topics of fake news are repurposed and covered by different fact-checking websites, we have included all of them in our data because content-wise each fact-checking website uses different claim to verify the title. For instance, article title "Vladimir Putin let lions loose in the streets of Russia to keep them indoors during COVID-19" and "Russia released 500 lions to ensure people would stay inside houses" both talk about the same topic but fact-checked by FactCrescendo and BOOM FactCheck.

## Conclusion and Future Work

In this work we presented a corpus of fact-checked articles on COVID-19. We collected news articles on COVID-19 from two data sources poynter and snopes and crawled the textual contents from the sources. An explanatory analysis is performed on the dataset to overview of data. We also discovered fact check articles which include linkages to social media. We manually annotated the fact-checked articles into 11 different categories. Finally, we built a machine learning-based classifier to detect misinformation about COVID-19. Our classifier performed with an F1-score of 0.76, which helped in the initial screening of the propagation of the misinformation at the time of the pandemic. The practical application of the classifier is to test the news article as false or true before doing the manual analysis. The limitation of this work is that due to the lack of polyglots, we provided the human annotated categories only for 3 languages.

One of the possible extension of this work could be analysis of the propagation of fake news across different social media platforms, for instance a case study with Twitter data is presented in (Shahi, Dirkson, and Majchrzak 2020). To optimise the search result of fake news on the search engine, publisher can also embed the class or category of fact checked news using schema markup(Shahi, Nandini, and Kumari 2019). Another possibility to broaden the work is to build a knowledge graph for fact check news so that the machine can easily process the data to answer user queries.

# Appendix

We have collected data from 92 distinct fact-checking website, and they have published the data in 86 different classes, we manually analyse and merge them into another class which has a similar meaning. Sometimes there is the only difference of a capital letter, for instance, false and False. Finally, we concluded all 86 classes into 23 unique class. The description of each class and its source(fact checking website) is mentioned in table 6.

For manual annotation of categories of fact-check articles, we have used a guideline which contains the definition of categories with an example. A complete description of 11 categories is shown in table 7.

# Acknowledgements

The work has been done under the Focus Group "COVID-19 Crisis" under RISE SMA project, which is funded by European Union's Horizon 2020 research and innovation programme under the Marie Skłodowska-Curie grant agreement No 823866.

# References

Augenstein, I.; Lioma, C.; Wang, D.; Lima, L. C.; Hansen, C.; Hansen, C.; and Simonsen, J. G. 2019. Multifc: A real-world multi-domain dataset for evidence-based fact checking of claims. *arXiv preprint arXiv:1909.03242*.

Cascella, M.; Rajnik, M.; Cuomo, A.; Dulebohn, S. C.; and Di Napoli, R. 2020. Features, evaluation and treatment coronavirus (covid-19). In *Statpearls [internet]*. StatPearls Publishing.

Lombard, M.; Snyder-Duch, J.; and Bracken, C. C. 2002. Content analysis in mass communication: Assessment and reporting of intercoder reliability. *Human communication research* 28(4):587–604.

Loper, E., and Bird, S. 2002. Nltk: the natural language toolkit. *arXiv preprint cs/0205028*.

Loria, S. 2018. textblob documentation. *Release 0.15* 2.

Mesquita, C. T.; Oliveira, A.; Seixas, F. L.; and Paes, A. 2020. Infodemia, fake news and medicine: Science and the quest for truth. *International Journal of Cardiovascular Sciences* (AHEAD).

Moriguchi, T.; Harii, N.; Goto, J.; Harada, D.; Sugawara, H.; Takamino, J.; Ueno, M.; Sakata, H.; Kondo, K.; Myose, N.; et al. 2020. A first case of meningitis/encephalitis associated with sars-coronavirus-2. *International Journal of Infectious Diseases*.

Organization, W. H., et al. 2020a. Coronavirus disease 2019 (covid-19): situation report, 72.

Organization, W. H., et al. 2020b. Novel coronavirus ( 2019-ncov): situation report, 3.

Orso, D.; Federici, N.; Copetti, R.; Vetrugno, L.; and Bove, T. 2020. Infodemic and the spread of fake news in the covid-19-era. *European Journal of Emergency Medicine*.

Parikh, S. B., and Atrey, P. K. 2018. Media-rich fake news detection: A survey. In *2018 IEEE Conference on Multimedia Information Processing and Retrieval (MIPR)*, 436–441. IEEE.

Pérez-Rosas, V.; Kleinberg, B.; Lefevre, A.; and Mihalcea, R. 2017. Automatic detection of fake news. *arXiv preprint arXiv:1708.07104*.

Poynter Institute. 2020. *The International Fact-Checking Network*.

Richardson, L. 2007. Beautiful soup documentation. *April*.

Rovetta, A., and Bhagavathula, A. S. 2020. Covid-19-related web search behaviors and infodemic attitudes in italy: Infodemiological study. *JMIR Public Health and Surveillance* 6(2):e19374.

Shahi, G. K.; Dirkson, A.; and Majchrzak, T. A. 2020. An exploratory study of covid-19 misinformation on twitter. *arXiv preprint arXiv:2005.05710*.

Shahi, G. K.; Nandini, D.; and Kumari, S. 2019. Inducing schema. org markup from natural language context. *Kalpa Publications in Computing* 10:38–42.

Shuyo, N. 2014. Language-detection library.

Snopes. 2020. Collections archive.

Tidy, J. 2020. Coronavirus: Facebook alters virus action after damning misinformation report.

Vlachos, A., and Riedel, S. 2014. Fact checking: Task definition and dataset construction. In *Proceedings of the ACL 2014 Workshop on Language Technologies and Computational Social Science*, 18–22.

Wang, W. Y. 2017. " liar, liar pants on fire": A new benchmark dataset for fake news detection. *arXiv preprint arXiv:1705.00648*.

Zarocostas, J. 2020. How to fight an infodemic. *The Lancet* 395(10225):676.

Zhou, X., and Zafarani, R. 2018. Fake news: A survey of research, detection methods, and opportunities. *arXiv preprint arXiv:1812.00315*.

Table 6: Definition of different type of fact-check articles

| Class | Definition | Source |
|---|---|---|
| No rating | The fact-checking entity decided not to apply any rating to this article. | www.factcheck.org |
| True | The rated statements are demonstrably true and no significant details are missing. | www.snopes.com |
| Mostly True | The primary elements of the rated statements are demonstrably true, however, there are minor errors, missing information or statements that need further clarification. | www.snopes.com |
| Partially True | The rated statements are partially correct but leave out important details, includes major errors or takes aspects out of context. | www.snopes.com |
| Mixture | The rated statements contain both significant true and significant false elements, such as exaggerations or false details. The available evidence behind the rated statements may also be evenly weighted in support of and against the claim. | www.snopes.com |
| Partially False | The rated statements are mostly false or not backed by evidence, but there is more than one element of truth. | www.animalpolitico.com |
| Mostly False | The primary elements of the rated statements are demonstrably false, however, there are minor details that are accurate. | www.politifact.com |
| False | The rated statements are demonstrably false. | www.healthfeedback.org |
| Four Pinocchios or Pants on Fire | The rated statements are demonstrably false and make a ridiculous claim, major exaggeration or make fear-mongering statements with the intent to provoke a panic reaction. | www.politifact.com |
| Misinformation or Misattributed | The rated statements seem to be in line with available evidence, but are used to reach erroneous conclusions, are misattributed or there is no consensus on what is the correct interpretation of the evidence. | www.colombiacheck.com |
| Misleading | The rated statements are backed by evidence, however, provided without necessary details or critical background knowledge and thus leaves the reader with a false understanding of reality. | www.healthfeedback.org |
| Unsupported or Unproven | The rated statements are not backed by reliable evidence and can neither be proven right nor wrong. Falsification of the rated statement would require difficult or impossible methods, and the claim is unverifiable. | www.dubawa.org |
| Manipulation | The rated statements contain claims that are beyond misleading or are based on methods that can be easily manipulated or framed in a manipulative way. | www.factcheck.kz |
| Out Of Context | The rated statements appear to be accurate, however, they were taken out of context, put into the wrong context, framed with misleading information, or used to create inaccurate connections. | www.leadstories.com |
| Pseudoscience | The rated statements are based on evidence that is not scientifically acceptable nor reliable. | www.ellinikahoaxes.gr |
| Clickbait | The rated statements were published as a headline that does not match the article's content nor scientific evidence and was formulated with excessive markup in order to create a usually unjustified overly emotional response that creates the most traffic/attention for the statement's creator. | www.raskrinkavanje.ba |
| Outdated | The rated statements were made based on evidence that used to be accepted but has now been proven inaccurate or irrelevant. | www.snopes.org |
| Labelled Satire | The rated statements were created as or based on content that, by the creator and a wider audience, was labelled as satire, whether this label is inaccurate or not. | www.snopes.com |

Table 6 – *Continued from previous page*

| Class | Definition | Source |
|---|---|---|
| Legend | The rated statements contain or are based on events that are so general or lack detail to an extent where those events may have happened to someone, somewhere, somewhen, and are therefore essentially neither provable nor unprovable. | www.snopes.com |
| Conspiracy Theory | The rated statements were made based on some evidence that is usually blown out of proportion, taken out of context and used to make wrong connections, mixed with ridiculous and unsupported claims and can mostly be proven wrong. The creator of the statement is not an expert but may give the impression to be an expert or have some form of authority, in order to manipulate recipients to believe the claims. | www.lemonde.fr |
| Explanatory | The rated content is not a statement but an explanation of a fact or statement. | www.agenciaocote.com |
| Scam | This rating does not evaluate the truth of a statement, instead indicated outlets or pages that are verified scams or describe verified scams. | www.snopes.com |
| Miscaptioned | The rated media content has been proven real/non-manipulated/non-edited but have been used misleadingly or is accompanied by false explanatory material describing the content's origin, context and/or meaning. | www.snopes.org |

Table 7: Definition of different type of fact-check articles

| Class | Definition | Example |
|---|---|---|
| Origins & Spread | The article discusses news covering theories, statements, research and facts regarding the origin and the spread of COVID-19 among individuals, local communities or the global population. | Was COVID-19 Found in Packages of Toilet Paper? |
| Prevention & Treatments | The article discusses news covering theories, statements, research and facts regarding possible prevention tactics or treatments for COVID-19, including news regarding a vaccine. | Will Sipping Water Every 15 Minutes Prevent a Coronavirus Infection? |
| International Response | The article discusses news covering the response of international governmental and administrative institutions to COVID-19. | Does Video Show Guns, Violence in the aftermath of the Coronavirus Outbreak in China? |
| Conspiracy Theories | The article discusses news covering popular conspiracy theories regarding e.g. the existence, origin, spread and effects of COVID-19 or individual or governmental involvement in its creation and spread. | Does George Soros Own a Lab That 'Developed' COVID-19? |
| Prophecies & Predictions | The article discusses news covering individual's or researchers' prophecies and predictions regarding aspects (e.g. spread) of COVID-19. | Did Nostradamus Predict the COVID-19 Pandemic? |
| Business & Industry | The article discusses news covering the impact of the COVID-19 pandemic on businesses and industries or their response to the global pandemic. | Is Starbucks Offering $100 Coupons During the COVID-19 Pandemic? |
| Humanitarian Response | The article discusses news covering the response of humanitarian organisations, NGOs or individuals with the aim of humanitarian aid to the COVID-19 pandemic. | While everybody is busy with coronavirus, authorities in Görlitz, in Germany, secretly brought asylum seekers into the city. |
| Other diseases | The article discusses news covering other diseases with reference to COVID-19, in regard to similarities or development and spread at the same time as the COVID-19 pandemic. | In China, there is now also an outbreak of the hantavirus, with a first recorded death. |
| Viral Post | The article discusses statements of viral social media posts (e.g. viral memes or rapidly shared posts) regarding COVID-19. | Central Park hospital tents housed thousands of abused children released from underground captivity. |
| Fear Mongering | The article discusses fear-mongering statements whose purpose appears to provoke of discriminatory reactions (e.g. racism, irreligion, homophobia) or panic. | Viral publication says a video shows a "gay party in Italy few weeks before COVID-19. |
| Media Coverage | The article which discuss how well/poorly media is reporting on the coronavirus situation. | We did a Q&A on the facts about the coronavirus. |